\documentclass{article}
\usepackage{epsf,emulateapj,apjfonts}

\lefthead{IRWIN \& SARAZIN}
\righthead{LMXBS IN X-RAY--FAINT ELLIPTICALS}
\slugcomment{Astrophysical Journal Letters, in press}

\begin{document}

\title{Low Mass X-ray Binaries As the Source of the Very Soft X-ray Emission
in the X-ray Faintest Early-Type Galaxies}

\author{Jimmy A. Irwin}
\affil{Department of Astronomy, University of Michigan, \\
Ann Arbor, MI 48109-1090 \\
E-mail: jai7e@astro.lsa.umich.edu}

\and

\author{Craig L. Sarazin}
\affil{Department of Astronomy, University of Virginia, \\
P. O. Box 3818, Charlottesville, VA 22903-0818 \\
E-mail: cls7i@virginia.edu}

\begin{abstract}
The X-ray emission from the faintest X-ray elliptical and S0 galaxies is
characterized by a hard $\sim$5 keV component, and a very soft
$\sim$0.2 keV component. The hard component has generally been regarded
as the integrated emission from low mass X-ray binaries (LMXBs), but the
origin of the soft component is unknown. We present evidence which
suggests that LMXBs also exhibit a soft component, which is responsible
for the very soft X-ray emission in the faintest early-type galaxies.
This soft component is present in two Galactic LMXBs which lie in directions
of low Galactic hydrogen column densities, and in LMXBs in the bulge of M31,
which comprise a majority of the X-ray emission in the bulge of that galaxy.
The X-ray spectral characteristics and X-ray--to--blue luminosity ratios of the
bulges of M31 and the Sa galaxy NGC~1291 are very similar to those of the
X-ray faintest early-type galaxies, indicating that LMXBs are responsible
for both soft and hard components in the latter. In addition, a low temperature
interstellar medium might be present in some X-ray faint galaxies.

\end{abstract}


\section{INTRODUCTION} \label{sec:intro}

Although abundant evidence exists that the X-ray emission in bright
early-type galaxies originates from a hot ($\sim$0.8 keV) interstellar medium,
the nature of the X-ray emission in X-ray faint (having a low X-ray--to--optical
luminosity ratio, $L_X/L_B$) early-type galaxies is far from certain.
{\it ROSAT} and {\it ASCA} spectra of several X-ray faint galaxies are
adequately fit by a two component thermal model in which one component has a
temperature of several keV, and the other a temperature of $\sim$0.2 keV
(Fabbiano, Kim, \& Trinchieri 1994; Pellegrini 1994; Kim et al.\ 1996).
The hard component has been attributed to the integrated emission from
low mass X-ray binaries (LMXBs); this component is seen in the bulge
of our own Galaxy.
X-ray binary stars (LMXBs and high mass X-ray binaries) are
believed to be the dominant source of X-ray emission in spiral galaxies
(Trinchieri \& Fabbiano 1985). Such a hard component appears to be present
in many elliptical and S0 galaxies observed with {\it ASCA}, even those
with high $L_X/L_B$ (Matsumoto et al.\ 1997).

The origin of the soft component has been the focus of considerable debate.
Emission from M star coronae, RS CVn binary systems, and supersoft sources
have been suggested as the source because of their soft X-ray properties,
but none of these sources appear to contribute appreciably to
the X-ray emission of the faintest early-type galaxies
(Pellegrini \& Fabbiano 1994; Irwin \& Sarazin 1997; hereafter IS97).
These sources are
either too faint to account for the emission, or possess X-ray spectral
characteristics which differ from the observed characteristics of the
emission from the X-ray faint galaxies.

In this {\it Letter} we propose that LMXBs are responsible for a majority of
the soft X-ray emission in addition to the hard emission. Such a soft
component is seen in LMXBs in the bulge of M31, and in two Galactic
LMXBs which lie in directions of low Galactic hydrogen column densities. The
X-ray luminosities and spectral characteristics of the X-ray faintest
early-type galaxies are consistent with what would be expected from
a collection of LMXBs. In some of the brighter X-ray faint galaxies, a
low temperature ($\sim$0.2 keV) ISM may also contribute appreciably to
the X-ray emission.

\section{X-RAY COLORS AND LUMINOSITIES}
\label{sec:colors_luminosities}

A sample of 61 early-type galaxies observed with the {\it ROSAT} PSPC was
used in this study. The selection criteria for the sample as well as the
data reduction are described in IS97. The
soft ($0.11-0.41$ keV; denoted `s' for soft) X-ray--to--optical luminosity
ratio, $L_X^s/L_B$, and the hard ($0.52-2.02$ keV; denoted `h' for hard)
X-ray--to--optical luminosity ratio, $L_X^h/L_B$, were calculated for
each galaxy in the sample. All galaxies with $\log(L^h_X/L_B) < 29.7$
(in units of ergs s$^{-1}$ $L_{B,\odot}^{-1}$) were labeled as X-ray faint
galaxies. Also calculated were two X-ray ``colors", C21 and C32, defined as
\begin{equation} \label{eq:c21}
{\rm C21} =
\frac{\rm counts~in~0.52-0.90~keV~band}{\rm counts~in~0.11-0.41~keV~band}
\, ,
\end{equation}
and
\begin{equation} \label{eq:c32}
{\rm C32} =
\frac{\rm counts~in~0.91-2.02~keV~band}{\rm counts~in~0.52-0.90~keV~band}
\, .
\end{equation}
X-ray colors are useful for extracting spectral information
from observations which contain too few counts to perform detailed
spectral modeling, as is often the case for the X-ray faintest galaxies.
Once calculated, the colors were corrected for foreground Galactic absorption,
which affects the C21 color significantly and the C32 color to a lesser extent.
To accomplish this, we determined the emitted colors (C21 and C32) and the
absorbed colors (C21$^*$ and C32$^*$) for an extensive grid of
Raymond-Smith (1977) thermal models with varying abundances
(ranging from 20\% to 100\% of solar) and temperatures
(ranging from 0.2 to 1.5 keV).
They were subject to different amounts of foreground absorption, ranging from
$N_H = 0.6$ to $6 \times 10^{20}$ cm$^{-2}$; the upper limit corresponds
to the upper limit allowed for inclusion in our sample.
Using XSPEC, these models were folded through the PSPC spectral response.
Empirically-determined functions were found which related C21 and C32 to
C21$^*$ and C32$^*$ as a function of $N_H$ (see IS97 for details).
This method was quite effective in
correcting the colors for absorption, producing only a 6\% rms error in C21
and a 2\% rms error in C32 when comparing the absorption-corrected colors to
the original unabsorbed colors derived from the spectral models.

It was found that the colors of the X-ray faint
galaxies were substantially different from those of the rest of the
galaxies in the sample. Whereas the X-ray brighter galaxies had colors
consistent with thermal emission from metal-enriched ($10\%-100\%$ of the
solar value) gas at temperatures
of $0.6-1.5$ keV, the X-ray faintest galaxies had very low values of C21,
the color which is sensitive to very soft X-ray emission. Although the
colors of some of the X-ray faintest galaxies were consistent with
thermal models with zero metallicity, such a model is not a physically
plausible one, since it would be expected that any interstellar medium
(ISM) in the galaxy would be polluted with some heavy elements
from Type Ia supernovae and/or normal stellar mass loss
(Fabbiano et al.\ 1994).
Furthermore, some of the X-ray faintest galaxies had colors which
were not even consistent with zero metallicity.

\centerline{\null}
\vskip2.65truein
\includegraphics{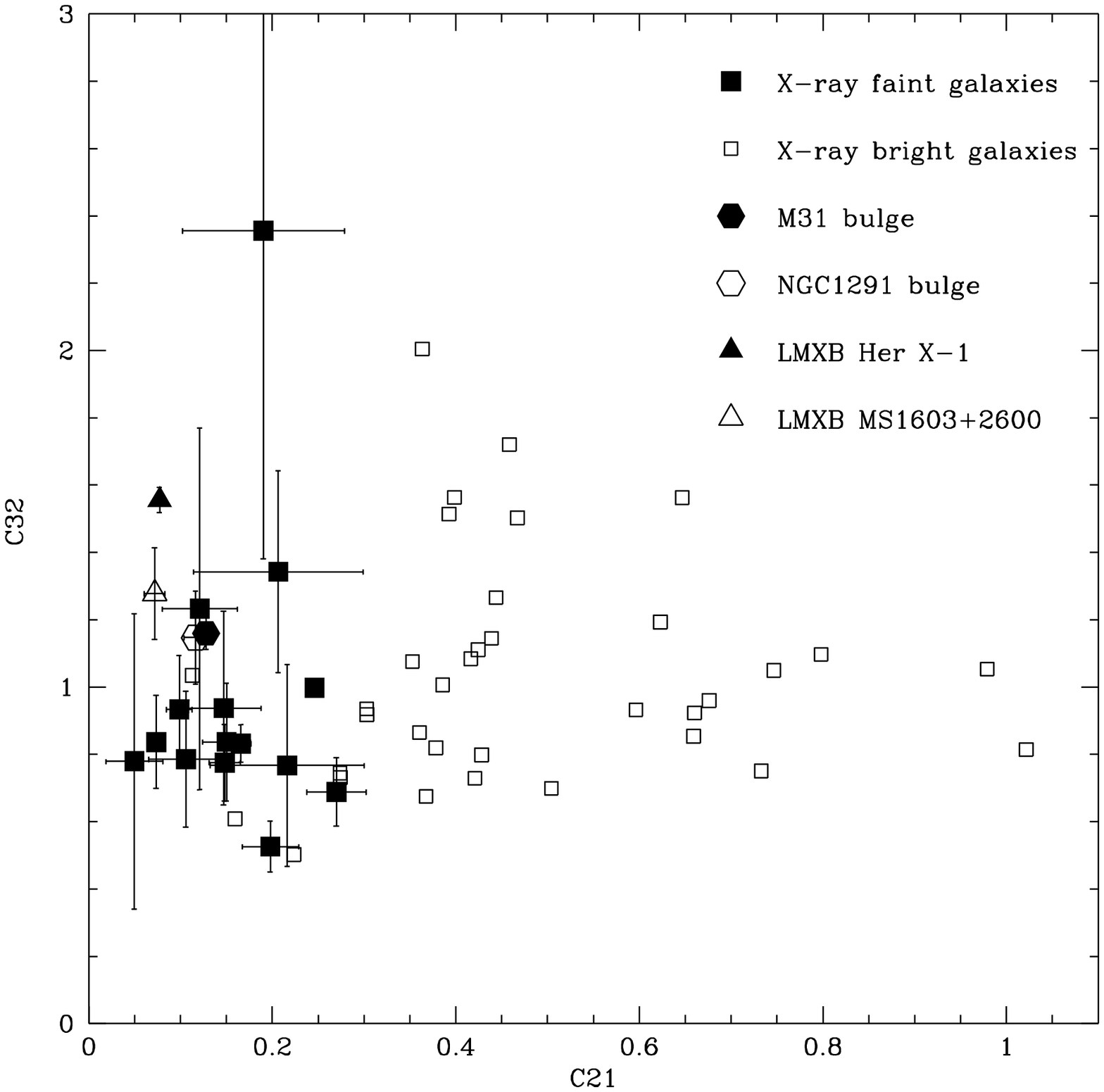}
\figcaption{
Plot of X-ray colors C21 vs.\ C32 for all galaxies in the sample with
detectable emission with $1\sigma$ error bars. X-ray faint early-type
galaxies (filled squares) clearly occupy a different region of color-color
space from their X-ray bright counterparts. The bulges of M31 and NGC~1291
(hexagons) and two Galactic LMXBs (triangles) have similarly soft C21 colors
as the X-ray faint early-type galaxies. Many of the X-ray faint
galaxies lie slightly below and to the right of the spiral bulges, possibly
due to the presence of warm ISM in addition to LMXBs.
\label{fig:colors}}

\vskip0.2truein

A summary of the average C21 and C32 values, as well as the observed range
of $\log(L^h_X/L_B)$ and $\log(L^s_X/L_B)$ values for the X-ray faintest
galaxies is given in Table~\ref{tab:average_values}. The errors on the
colors are the standard deviations for the sample rather than the statistical
errors, in order to better represent the scatter of the property.
The individual X-ray colors and X-ray--to--optical luminosity ratios for the
galaxies classified as X-ray faint are plotted in Figure~\ref{fig:colors}
and Figure~\ref{fig:lxlb}, respectively.

\section{COMPARISON TO SPIRAL BULGES AND GALACTIC LMXBS}
\label{sec:spiral_bulges}

Published estimates for the stellar contribution to the X-ray emission 
of early-type galaxies vary by more than a factor of 10
(Forman et al.\ 1985; Canizares, Fabbiano, \& Trinchieri 1987),
so it is not certain at what level stellar X-ray emission should become
important. If stellar emission is important in the X-ray faintest
early-type galaxies, then we might expect to see the same spectral
characteristics in the bulges of spiral galaxies, whose old stellar
population is similar to that of early-type galaxies. The bulge of
M31 is a suitable candidate against which to test this hypothesis, since it
is close enough that individual X-ray point sources can be resolved
with {\it ROSAT}.

\centerline{\null}
\vskip2.65truein
\includegraphics{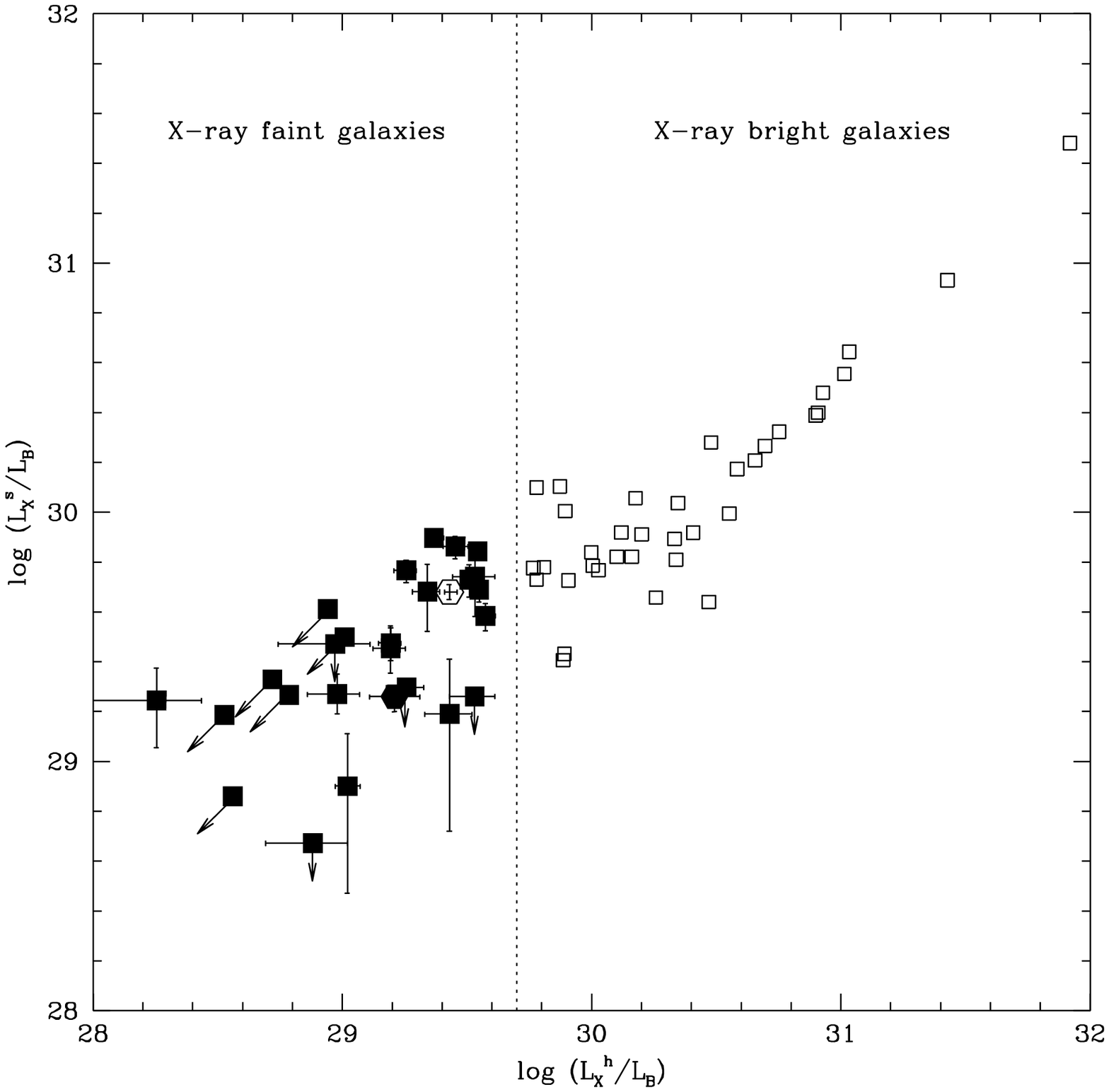}
\figcaption{
Plot of $\log({L_X^s/L_B})$ vs.\ $\log({L_X^h/L_B})$ for all galaxies in the
survey. Symbols with one-sided vertical arrows were detected in the hard
band only, and symbols with one-sided slanted arrows were not detected in
either band. The X-ray faint galaxies show a roughly linear relation between
$\log({L_X^s/L_B})$ and $\log({L_X^h/L_B})$. The bulge of M31 (filled
hexagon) and the bulge of NGC~1291 (open hexagon) lie within the observed
range of values for both quantities. The non-constant value for both ratios
among galaxies whose X-ray emission appears to be primarily stellar
suggests that the relative proportion of LMXBs varies from galaxy to
galaxy.
\label{fig:lxlb}}

\vskip0.2truein

Supper et al.\ (1997) tabulated the count rates of 22 point sources
found within $5^{\prime}$ of the center of the bulge of M31 in the same
three {\it ROSAT} energy bands which we used to calculate our X-ray
colors. We summed the count rates for the 22 sources in each of the three
energy bands and calculated the integrated X-ray colors. The colors were 
corrected for Galactic absorption towards the direction of M31
($6.73 \times 10^{20}$ cm$^{-2}$; Stark et al.\ 1992). The colors of the
sum of the sources [(C21, C32$)=(0.08, 1.15)$]
were remarkably similar to those of the X-ray faintest early-type galaxies.
Supper et al.\ (1997) found that
these 22 sources comprise 75\% of the total X-ray emission from within
$5^{\prime}$ of the bulge. Using {\it ROSAT} PSPC archival data of M31,
we found that within a $5^{\prime}$ extraction radius the
colors of the total emission from the bulge (resolved and unresolved)
were (C21, C32$)=(0.13, 1.16)$ (see Figure~\ref{fig:colors}). Thus, the
remaining 25\% of unresolved flux only moderately increases the C21 color,
while the C32 color is unchanged.

Also calculated were the $\log(L_X^h/L_B)$ and $\log(L_X^s/L_B)$ values
of the inner 10$^{\prime}$ of the bulge of M31 (which corresponds to
the effective half-light radius of the bulge). The results are shown
in Table~\ref{tab:average_values} and Figure~\ref{fig:lxlb}. Both
X-ray--to--optical luminosity ratios fall within the observed ranges found
for the X-ray faintest early-type galaxies.

\begin{table*}[tbp]
\caption[X-ray Properties]{}
\label{tab:average_values}
\begin{center}
\begin{tabular}{lcccc}
\multicolumn{5}{c}{\sc X-ray Colors and X-ray--to--Optical Luminosity Ratios}
\cr
\tableline \tableline
Sample& C21 & C32 & $\log(L_X^h/L_B)$ & $\log(L_X^s/L_B)$ \\
&&& (ergs s$^{-1} L_{B,\odot}^{-1}$) & (ergs s$^{-1} L_{B,\odot}^{-1}$) \\
\tableline
X-ray faint galaxies & $0.16\pm0.06 $& $0.99\pm0.41$
& $<$29.7\phn & $<$29.9\phn \\
M31 bulge (integrated) & $0.13\pm0.01$ & $1.16\pm0.05$ &
\phantom{$>$}29.21 & \phantom{$>$}29.26 \\
M31 bulge (point sources only) & $0.08\pm0.01$ & $1.15\pm0.07$ & \ldots &
\ldots \\
NGC~1291 bulge & $0.12\pm0.01$ & $1.15\pm0.14$ &
\phantom{$>$}29.43 & \phantom{$>$}29.68 \\ 
Her X-1 & $0.08\pm0.02$ & $1.56\pm0.04$ & \ldots & \ldots \\
MS~1603+2600 & $0.07\pm0.01$ & $1.28\pm0.14$ & \ldots & \ldots \\
\tableline
\end{tabular}
\end{center}
\end{table*}

%Three of the point sources in the bulge of M31 show no emission above
%0.4 keV, and are most likely supersoft X-ray sources, as pointed out by
%Supper et al.\ (1997). These sources are faint relative to the other
%sources and do not significantly affect the X-ray colors. The other 19 sources
The point sources in the bulge of M31
have X-ray luminosities in the range $10^{36}-10^{38}$ ergs s$^{-1}$.
Since the brightest low mass X-ray binaries in our Galaxy have X-ray
luminosities on this order, it is likely that a majority of these sources
are LMXBs.
Individually, all but two of these point sources exhibit a very soft C21 color,
similar to those seen in the X-ray faintest early-type galaxies;
the soft C21 color seen in the integrated emission from the bulge of
M31 is not due to one or two extremely bright, extremely soft sources.
Since the majority of the X-ray emission in the bulge of M31 is from
LMXBs, and these LMXBs have X-ray characteristics similar to those
of the X-ray faintest early-type galaxies, this strongly suggests that the
bulk of the X-ray emission in the latter results from LMXBs also. These
galaxies might also contain small amounts of warm ISM; this will be discussed
in \S~\ref{sec:discussion}.

To confirm these results, we also determined the X-ray colors and
X-ray--to--optical luminosity ratios for the bulge of the bright Sa galaxy
NGC~1291. Although this galaxy is too far away for the X-ray emission to be
resolved with {\it ROSAT},
nearly identical values for the colors were found for NGC~1291 as
M31 (see Table~\ref{tab:average_values} and Figure~\ref{fig:colors}).
Somewhat higher $L_X/L_B$ values were found for NGC~1291, but still within
the observed range among X-ray faint elliptical and S0 galaxies
(Figure~\ref{fig:lxlb}).

It would be very useful to search for this soft component in Galactic LMXBs.
Unfortunately, most Galactic LMXBs lie near the Galactic
plane, where high hydrogen column densities completely absorb very
soft X-rays. Two Galactic LMXBs which lie in directions of low column
densities are Her X-1 ($N_H=1.73\times10^{20}$ cm$^{-2}$) and MS~1603+2600
($N_H=4.57\times10^{20}$ cm$^{-2}$). Archival PSPC data of these two
objects were extracted, and once corrected for absorption, these
two LMXBs had X-ray colors similar to the bulges of M31 and NGC~1291,
and the X-ray faint early-type galaxies (see Table~\ref{tab:average_values}
and Figure~\ref{fig:colors}). Specifically, very soft C21 colors were
found for both Galactic LMXBs, indicative of very soft X-ray emission.

On the other hand, a very soft C21 color was not found for
two globular cluster LMXBs, or LMC X-2 in the Large Magellanic Cloud.
Verbunt et al.\ (1995) tabulated the count rates in our three bands for LMXBs
in the globular clusters NGC~1851 and NGC~7078 (M15). After correcting
the count rates for Galactic absorption, values of
(C21, C32$)=(0.30, 2.37)$ and (C21, C32$)=(0.23, 3.28)$ were obtained for
these two cluster LMXBs. In addition, spectral modeling of the archival data
of LMC X-2 showed the emission from this LMXB was described adequately by a
two component bremsstrahlung model. The colors of this model were
(C21, C32$)=(0.24, 1.57)$. These colors are significantly higher than the
colors of the LMXBs in the bulge of M31, Her X-1, and MS~1603+2600.

It is possible that the soft X-ray component in LMXBs depends on metallicity.
The two globular cluster LMXBs and LMC X-2 are found in environments with a
substantially lower metallicity than Her X-1, MS~1603+2600, and the
M31 bulge LMXBs.
The globular cluster NGC~7078 has a particularly low
metallicity (1\% solar; Adams et al.\ 1983) and the LMXB in this cluster
has the highest C32 value of the LMXBs studied here. The Large
Magellanic Cloud has a metallicity intermediate between the globular
clusters and the bulges of M31 and NGC~1291, and its LMXB has a C32 color
intermediate between those two systems.
Given this possible dependence of the X-ray spectral properties on metallicity,
it is probably best to use the M31 bulge LMXBs as a template for the
X-ray emission expected from LMXBs in early-type galaxies, since the
metallicities of the stellar populations are comparable in these two systems.

\section{DISCUSSION} \label{sec:discussion}

If the majority of the X-ray emission from X-ray faint early-type galaxies
is from stellar sources, this would imply that most of the
gas lost from stars has not been retained by the galaxy.
Hydrodynamical simulations of the gas in elliptical galaxies
(Loewenstein \& Mathews 1987;
David, Forman, \& Jones 1991;
Ciotti et al.\ 1991)
have shown that some galaxies may be experiencing a wind phase in which
Type Ia supernovae drive most of the gas lost from stars out of the galaxy.
Alternatively, the gas may have been lost by ram pressure stripping or
other external effects.
In this scenario, the X-ray bright galaxies
have retained most of their ISM while the X-ray faint galaxies have lost
most of their ISM.
The X-ray faint galaxies in our sample generally tend to have rather low
stellar velocity dispersions, on the order of 200 km s$^{-1}$.
As such, they have smaller gravitational potential wells than their X-ray
bright counterparts, and this would make it easier to remove the
gas from them.

Examination of Figure~\ref{fig:colors} reveals that although the colors of
the majority of the X-ray faint galaxies are consistent with the colors of
M31 and NGC~1291 at the $1\sigma$ level, many of the galaxies have slightly
lower C32 values than the spiral bulges. This could be due to the presence
of small amounts of warm ISM in these galaxies.
For example, if a Raymond-Smith thermal model with a temperature of
0.2 keV and a metallicity of 50\% solar is folded through the spectral
response of the PSPC, the X-ray colors are (C21, C32$)=(0.21, 0.20)$.
When the emission from
this component is mixed in the right proportion with the emission expected
from LMXBs, the colors of the X-ray faint galaxies are more accurately
reproduced. For example, if 85\% of the X-ray emission (by counts) is
from LMXBs with the colors of those in the bulge of M31
[(C21, C32$)=(0.08, 1.15)$] and 15\% from a 0.2 keV ISM
[(C21, C32$)=(0.21, 0.20)$], the resulting emission would have colors
of (C21, C32$)=(0.10, 0.86)$, in good agreement with what is observed
for X-ray faint galaxies.

The spread in colors among X-ray faint galaxies could be explained
by varying proportions of ISM emission. As the percentage of the X-ray emission
attributable to the ISM is increased, the galaxy would move down and to the
right on the color-color plot (Figure~\ref{fig:colors}).
The galaxies with the highest percentage
of ISM emission would also be expected to have the highest $L_X/L_B$ values.
In fact, the 5 most X-ray luminous galaxies classified as X-ray faint
lie furthest below and to the right of the spiral bulges.  In these galaxies,
the ISM may be responsible for as much as a quarter of the total X-ray emission.

Despite the fact that most of the X-ray emission in X-ray faint galaxies
appears to be stellar in nature, Figure~\ref{fig:lxlb} illustrates that there
does not seem to be a simple linear relation between the X-ray and optical
luminosities.
If there were, one would expect the X-ray faintest galaxies to cluster
at a fixed point in this Figure, given by the values of
$L_X^h/L_B$ and $L_X^s/L_B$ for purely stellar emission. However,
a range of $L_X/L_B$ values are present for galaxies whose emission is
suspected to be primarily stellar. The bulge of NGC~1291 has a $L_X^h/L_B$
value 1.7 times larger than that of M31. Although it is possible that NGC~1291
has an additional X-ray component not present in M31, this component would
have to have X-ray colors which were identical to the LMXB component,
since M31 and NGC~1291 have the same X-ray colors. This seems unlikely.

Figure~\ref{fig:lxlb} shows that there are 5 galaxies with upper limits on
$L_X^h/L_B$ which are at least a factor of 3 less than that of M31 (whose
X-ray emission is at least 75\% stellar; see \S~\ref{sec:spiral_bulges}).
Thus, at least for low optical luminosities, the X-ray emission does
not scale linearly with the optical luminosity.
The fact that the X-ray colors don't vary as the X-ray--to--optical
luminosity ratio varies (or equivalently, that $L_X^h/L_B$ and $L_X^s/L_B$
vary approximately linearly with one another) suggests
that there is one dominant X-ray emission mechanism at work here, but
that mechanism is stronger in some galaxies than others at a given optical
luminosity.
This variation would require that the relative number of
LMXBs vary from galaxy to galaxy (see IS97).

\section{CONCLUSIONS} \label{sec:conclusions}

We propose that the very soft X-ray emission seen in X-ray faint elliptical
and S0 galaxies results from LMXBs. The X-ray spectral properties of
these galaxies are very similar to those of LMXBs in the bulges of M31
and NGC~1291, and the Galactic LMXBs Her X-1 and MS~1603+2600. Furthermore,
LMXBs are luminous and numerous enough to account for the measured X-ray
luminosities in these galaxies. Some warm ISM may also be present in these
systems in small amounts.

Since LMXBs are luminous X-ray sources, this model predicts that the
soft X-ray component in X-ray faint ellipticals will be resolvable into
discrete sources given better spatial resolution and sensitivity than
is possible with the {\it ROSAT} telescope.
Given current estimates of their sensitivity, we find that the
backside-illuminated CCD chips of the ACIS instrument on {\it AXAF}
will have
sufficient soft X-ray sensitivity, spectral resolution, and spatial resolution
to detect and resolve individual LMXBs in early-type galaxies at the distance
of the Virgo cluster or closer.
%The HRI instrument has sufficient sensitivity and spatial resolution,
%but probably will lack sufficient spectral resolution to separately image
%the soft X-ray component in these systems.

\acknowledgments

We thank Michael Loewenstein for many very useful comments and suggestions.
We also thank Steve Balbus, Tim Kallman, Bob O'Connell, Richard Mushotzky,
and Jean Swank for useful discussions.
This research has made use of data obtained through the High Energy
Astrophysics Science Archive Research Center Online Service, provided
by the NASA/Goddard Space Flight Center.
J. A. I. and C. L. S. were supported in part by NASA ROSAT grant NAG 5--3308,
and ASCA grant NAG 5-2526.
C. L. S. was also supported by NASA Astrophysical Theory Program grant 5-3057.
J. A. I. was supported by the Achievement Rewards for College Scientists
Fellowship, Metropolitan Washington Chapter and NASA grant NAG 5-3247.

\end{document}